%% file: main.tex
\documentclass{article}

\usepackage{microtype}
\usepackage{graphicx}
\usepackage{subfigure}
\usepackage{booktabs} 

\usepackage{hyperref}

\usepackage[accepted]{icml2025}

\usepackage{amsmath}
\usepackage{amssymb}
\usepackage{mathtools}
\usepackage{amsthm}
\usepackage{float}
\usepackage{threeparttable}
\usepackage{placeins}
\usepackage{listings}
\usepackage{ebproof}
\usepackage{url}
\usepackage{enumitem}

\usepackage{dashbox}
\usepackage[
    n,
    operators,
    advantage,
    sets,
    adversary,
    landau,
    probability,
    notions,    
    ff,
    mm,
    primitives,
    events,
    complexity,
    asymptotics,
    keys]{cryptocode}

\usepackage[capitalize,noabbrev]{cleveref}

\theoremstyle{plain}
\newtheorem{theorem}{Theorem}[section]

\theoremstyle{definition}
\newtheorem{definition}[theorem]{Definition}

\theoremstyle{remark}

\usepackage[textsize=tiny]{todonotes}

\definecolor{red}{RGB}{220,40,30}
\definecolor{BrightOrange}{RGB}{255,140,0}
\definecolor{Gold}{RGB}{255,200,0}
\definecolor{Emerald}{RGB}{0,160,80}
\definecolor{Cobalt}{RGB}{0,120,215}
\definecolor{Sapphire}{RGB}{90,60,170}
\definecolor{Amethyst}{RGB}{160,80,200}

\begin{document}

\twocolumn[
\icmltitle{Towards Provable (In)Secure Model Weight Release Schemes}



\icmlsetsymbol{equal}{*}

\begin{icmlauthorlist}
\icmlauthor{Xin Yang}{zju}
\icmlauthor{Bintao Tang}{tongji}
\icmlauthor{Yuhao Wang}{tongji}
\icmlauthor{Zimo Ji}{ust}
\icmlauthor{Terry Jingchen Zhang}{eth}
\icmlauthor{Wenyuan Jiang}{eth}
\end{icmlauthorlist}

\icmlaffiliation{zju}{Polytechnic Institute, Zhejiang University, Zhejiang, China}
\icmlaffiliation{tongji}{School of Software Engineering, Tongji University, Shanghai, China}
\icmlaffiliation{ust}{Department of Computer Science and Engineering, Hong Kong University of Science and Technology, Clear Water Bay, Hong Kong}
\icmlaffiliation{eth}{ETH Zurich, Zurich, Switzerland}

\icmlcorrespondingauthor{Zimo Ji}{zjiag@connect.ust.hk}
\icmlcorrespondingauthor{Wenyuan Jiang}{wenyjiang@ethz.ch}

\icmlkeywords{Machine Learning, ICML}

\vskip 0.3in
]



\printAffiliationsAndNotice{}  

\setlist[itemize]{noitemsep, topsep=0pt}

\input{parts/0-Abstract}
\input{parts/1-Introduction}
\input{parts/2-RelatedWork}

\input{parts/3-SecurityDefinitions}

\input{parts/4-TaylerAttack}

\input{parts/5-FutureWork}

\input{parts/6-Conclusion}

\bibliography{refs}
\bibliographystyle{icml2025}

\newpage
\appendix
\input{parts/7-Appendix}

\end{document}

%% file: parts/0-Abstract.tex
\begin{abstract}
Recent secure weight release schemes claim to enable open-source model distribution while protecting model ownership and preventing misuse. 
However, these approaches lack rigorous security foundations and provide only informal security guarantees. 
Inspired by established works in cryptography, we formalize the security of weight release schemes by introducing several concrete security definitions. 
We then demonstrate our definition's utility through a case study of TaylorMLP, a prominent secure weight release scheme.
Our analysis reveals vulnerabilities that allow parameter extraction thus showing that TaylorMLP fails to achieve its informal security goals. 
We hope this work will advocate for rigorous research at the intersection of machine learning and security communities and provide a blueprint for how future weight release schemes should be designed and evaluated.
\end{abstract}

%% file: parts/1-Introduction.tex
\section{Introduction}
\label{sec:intro}

Deep learning models, especially large language models (LLMs), pose unique challenges for model weight release, the practice of sharing a model’s learned parameters with users.  
Providers typically choose between closed API hosting, which preserves developer control but forces users to expose private data to the service \cite{achiam2023gpt}, and fully open-sourcing the weights, which protects user privacy but cedes ownership and control of the model \cite{touvron2023llama}.  
This creates a dilemma: developers risk unauthorized extraction or repurposing of their proprietary weights if they release them openly, yet users face privacy and availability concerns when models are accessible only via APIs.

To address this, secure weight release schemes aim to enable local or offline inference without exposing raw weights, thereby balancing utility and protection.  
An ideal scheme would preserve accuracy and inference performance for legitimate users while making it difficult for an adversary to recover the original weights or to perform large-scale fine-tuning abuse.  
Such guarantees are crucial because model weights represent high-value intellectual property, often requiring millions of dollars in training investment \cite{refael2024slip}.

\begin{figure}[htbp]
    \centering
    \includegraphics[width=0.5\textwidth]{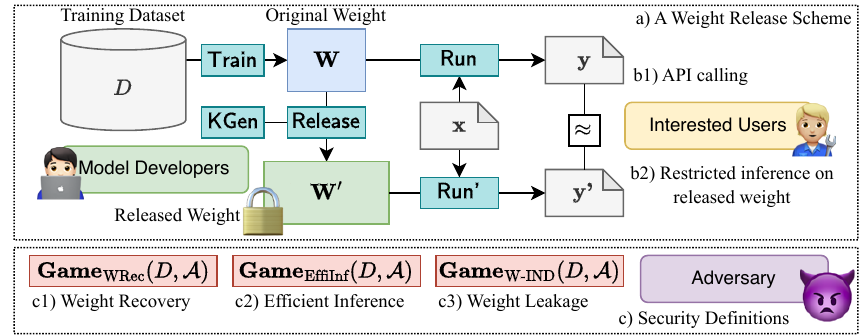}
    \vspace{-0.7cm}
    \caption{Overview of weight release schemes and our proposed security definitions. 
    (a) Model developers transform original weights into a restricted released model;
    (b) users either call an API (b1) or run local inference on the restricted weights (b2);
    (c) proposed game‐based security definitions: weight recovery (c1), efficient inference (c2) and weight leakage (c3).
    }
    \label{fig:mainfig}
\end{figure}

However, current weight release schemes like TaylorMLP \cite{wang-etal-2024-taylor} lack a formal security foundation.  Their security claims rest on informal hardness assumptions or empirical observations without reductions to well-studied hardness assumptions.  
Consequently, it remains unclear under what adversarial models and assumptions weight confidentiality truly holds, or how one might systematically evaluate a scheme’s (in)security.
Indeed, recent work shows that fine-tuning can often strip out empirical safeguards, hinting that ad hoc measures may be brittle  \cite{tamirisa2024tamper}.

In this paper, we fill this gap by introducing a formal framework for model weight release security, as outlined in Figure \ref{fig:mainfig}.
We define security properties, establish relationships among them, and show how they can guide scheme design. 
We then present a case study of TaylorMLP: we mount a parameter extraction attack that successfully recovers original weights, demonstrating that TaylorMLP falls short of its informal goals.
Finally, we distill our insights into a blueprint for future schemes, outlining design principles and evaluation criteria to achieve provable security guarantees.

Our contribution in this paper is threefold.
\begin{itemize}
    \item \textbf{Formal security properties of weight release schemes.} We defined several formal security properties of weight release schemes and proved some of their relations under this context.
    \item \textbf{A case study against TaylorMLP.} We performed an analysis on TaylorMLP and demonstrated that it is vulnerable to our parameter extraction attack, thus failing to deliver its claimed security goals.
    \item \textbf{A blueprint for future weight release schemes.} We provide a blueprint for how future weight release schemes should be designed and evaluated.
\end{itemize}

%% file: parts/2-RelatedWork.tex
\section{Related Work}
\label{sec:relatedwork}

\paragraph{Open-Source Models}
Open-source LLMs like LLaMA 2 \cite{touvron2023llama}, Qwen \cite{bai2023qwen}, and DeepSeek \cite{liu2024deepseek} have democratized access by releasing full model weights under permissive licenses. 
While this fosters innovation and reproducibility, it also hands over complete control of the models to downstream users. 
Without embedded technical safeguards, these releases rely solely on legal agreements to prevent misuse.

\paragraph{Privacy-Aware Model Inference}
Cryptographic protocols such as MPC and homomorphic encryption enable inference without exposing either model weights or user inputs, as demonstrated by BOLT \cite{pang2024bolt}, BumbleBee \cite{lu2023bumblebee}, and PUMA \cite{dong2023puma}. 
These systems offer provable security under standard assumptions but incur latency and resource costs orders of magnitude higher than native inference. 
Such overheads currently preclude their use for real-time or large-scale LLM deployment.

\paragraph{Weight Release Schemes}
TaylorMLP \cite{wang-etal-2024-taylor} publishes truncated Taylor-series coefficients of MLP weight matrices, allowing exact inference while obscuring the original parameters and introducing controllable throttling. 
Hardware obfuscation approaches embed DNNs in locked circuits that only function under specific configurations, preventing direct weight extraction \cite{goldstein2021preventing}. 
Its security is based on the presumed hardness of inverting the transformation, without formal reductions to established cryptographic problems.

\paragraph{Model Tracing}
Watermarking and fingerprinting embed hidden signals into model weights or outputs to detect unauthorized use. 
CoTGuard \cite{wen2025cotguard} and Double-\emph{i} Watermark \cite{li2024double} insert secret triggers into reasoning traces or fine-tuning, while Mark Your LLM \cite{xu2025mark} and ProFLingo \cite{jin2024proflingo} employ backdoor-based or query-based fingerprints. 
These techniques demonstrate empirical robustness against benign fine-tuning but lack strong guarantees against adaptive adversaries.

\paragraph{Model Abuse and Misuse Risks}
Open releases can be fine-tuned with minimal poisoned data to remove safety mitigations and repurpose models for harmful tasks. 
Surveys by Yan \emph{et al.} \cite{yan2024protecting} and Li \emph{et al.} \cite{li2023privacy} document how such attacks degrade alignment and enable sensitive data extraction. 
These risks motivate technical controls that enforce provable security properties in released models.


%% file: parts/3-SecurityDefinitions.tex
\section{Security of Weight Release Schemes}
\label{sec:security}


In this section, we first formally define the syntax of a weight release scheme.
Then we give several security properties for weight release schemes with formal definitions.
We also discuss the relations of these security properties and possible hardness assumptions under the context of weight release schemes.

\subsection{Weight release schemes}

Consider a deep learning task where training the model on dataset $D \in \mathcal{D}$ gives the model weight $W \in \mathcal{W}$. 
The model is expected to take $x \in \mathcal{X}$ as the input and output $y \in \mathcal{Y}$ when running inference on weight $W$. 
We then define the syntax of a weight release scheme as follows.

\paragraph{Syntax.}\label{sec:security-syntax} A weight release scheme is a tuple of 5 probabilistic algorithms $\Sigma = (\textsf{Train}, \textsf{Run}, \textsf{KGen}, \textsf{Release}, \textsf{Run'})$, where
\begin{itemize}
    \item $\textsf{Train} : \mathcal{D} \rightarrow \mathcal{W}$ abstracts the original training process on dataset $D \in \mathcal{D}$ and produces a model weight $W \in \mathcal{W}$.
    
    \item $\textsf{Run} : \mathcal{W} \times \mathcal{X} \rightarrow \mathcal{Y}$ abstracts the original inference process which runs $W \in \mathcal{W}$ on input $x \in \mathcal{X}$ and outputs $y \in \mathcal{Y}$.
    
    \item $\textsf{KGen} : \bot \rightarrow \mathcal{K}_\sk \times \mathcal{K}_\pk$ is similar to the public key cryptography case, which generates a pair of key $(\sk, \pk)$ for future use.
    There are many cases where the weight release scheme requires no key, and in this case $\mathcal{K}_\sk $ and $\mathcal{K}_\pk$ can both be $\emptyset$.
    
    \item $\textsf{Release} : \mathcal{K}_\sk \times \mathcal{W} \rightarrow \mathcal{W}'$ is the core of the weight release scheme which transforms the original weight $W \in \mathcal{W}$ to the released version $W' \in \mathcal{W}'$ under some private key $\sk$. 
    Note that depending on the security requirements, $\mathcal{W}'$ is generally not the same as $\mathcal{W}$, but there are cases like hidden watermarking where $\mathcal{W}' = \mathcal{W}$.
    
    \item $\textsf{Run'} : \mathcal{K}_\pk \times \mathcal{W}' \times \mathcal{X} \rightarrow \mathcal{Y}$ models the user with the public key $\pk$ inferencing the released version of weight $W' \in \mathcal{W}'$ on input $x \in \mathcal{X}$ and outputs $y \in \mathcal{Y}$.
\end{itemize}
Note that different from standard cryptographic practice where security parameters are included in the $\textsf{KGen}$ parameter, the security parameters for a specific weight release scheme are often fixed and implicitly determined by the scheme (e.g., model weight size). 

\paragraph{Game-based definition.} We use game-based definitions for formally specifying the properties of weight release schemes through interactive experiments in this section, following standard cryptographic practice for defining security properties of protocols\cite{bellare2004code,shoup2004sequences}. 
A game is a probabilistic experiment, often written as a procedure, that captures the essential behavior of the system whose output indicates the result of the experiment. 
In our context, games allow us to rigorously compare the behavior of original and released model weights while accounting for the inherent randomness in neural network training and inference processes.

\paragraph{Correctness.} Informally, a weight release scheme is correct if inferencing the released version of weight leads to \textit{same} results as inferencing the original weight.
Due to the intrinsic randomness in the inference process, defining \textit{same} is not like defining \textit{equality}.
Here we consider $\textsf{Dist}(a, b)$ as a distance function measuring how different $a, b \in \mathcal{S}$ are where $\mathcal{S}$ is the relevant domain for our definitions.
We also consider $\textsf{Same}(a, b)$ as $\textsf{Dist}(a, b) \leq \epsilon$ for some practically meaningful threshold $\delta$. 
For example, $\textsf{Dist}(W_a, W_b)$ for $W_a, W_b \in \mathcal{W}$ can be defined as $\left \| W_a - W_b \right \|_p $ where $\left \| \cdot \right \|_p$ is the matrix p-norm, and $\textsf{Same}(W_a, W_b)$ can be defined using $\delta = 10^{-3}$ for some specific model types.
Then correctness is defined as follows.

\begin{definition}[Correctness of a weight release scheme]
We define the following game.
\begin{pcvstack}[center,boxed]
\procedureblock[linenumbering]{$\textbf{Game}_\text{Correctness}(D)$}{
(\pk, \sk) \sample \textsf{KGen}() \\
W \sample \textsf{Train}(D) \\
W' \sample \textsf{Release}(\sk, W) \\
x \sample \mathcal{X} \\
y \sample \textsf{Run}(W, x) \\
y' \sample \textsf{Run'}(\pk, W', x) \\
\pcreturn \textsf{Same}(y', y)
}
\end{pcvstack}
Then a weight release scheme $\Sigma$ with $\Sigma = (\textsf{Train}, \textsf{Run}, \textsf{KGen}, \textsf{Release}, \textsf{Run'})$ is correct if 
\begin{align}
    \Pr[\textbf{Game}_\text{Correctness}(D) \Rightarrow 1] = 1
\end{align}
for given $D \in \mathcal{D}$.

\end{definition}


\subsection{Security properties}

Informally, a secure weight release scheme aims to 1) protect the model ownership of developers while 2) allow users to perform inference on the released model weight with \textit{reasonalble} efficiency.
Additional security goals include preventing abuse and further unintended modification of the released model. 
To formally define these security properties, we choose several typical security goals and formulate them into game-based definitions.
Each of the following security goals starts with an example scenario followed by a game and description of the adversary's abilities and constraints.

\paragraph{Preventing weight recovery.}
One natural security goal for secure weight release schemes is to prevent any adversary from recovering the original weight.
This goal can be captured by the following game for weight release scheme $\Sigma$ on a given training dataset $D \in \mathcal{D}$.
\begin{pcvstack}[center,boxed]
\procedureblock[linenumbering]{$\textbf{Game}_\text{WRec}(D, \mathcal{A})$}{
W \sample\textsf{Train}(D) \\
(\pk, \sk) \sample \textsf{KGen} () \\
W' \sample \textsf{Release}(\sk, W) \\
W^* \sample \adv(\pk, W') \\
\pcreturn \textsf{Same}(W^*, W)
}
\end{pcvstack}
The advantage of an adversary $\adv$ in this game is defined as
\begin{align}
    \advantage{\text{WRec}}{\Sigma}[(D, \adv)] = \Pr[\textbf{Game}_\text{WRec}(D, \adv) \Rightarrow 1]
\end{align}
\begin{definition}
Then for all efficient adversary $\adv$ on some dataset $D \in \mathcal{D}$, if 
\begin{align}
    \advantage{\text{WRec}}{\Sigma}[(D, \adv)] \leq \epsilon
\end{align}
for some negligible $\epsilon$ with regard to the security paramter of the scheme $\Sigma$, then the weight release scheme $\Sigma$ is considered to be $(D, \epsilon)$-weight-recovery-secure (WRec-secure).
\end{definition}
Note that since weight recovery is a very strong attack even for computational unbounded adversaries, because in $\textbf{Game}_\text{WRec}(D, \adv)$, there is randomness in training and $D$ is not given to $\adv$ thus making it intuitively infeasible for $\adv$ to recover the exact weight.
Therefore, security property aiming at preventing this attack is relatively weak compared with other security properties defined in later parts of this section.

Only ensuring $(D, \epsilon)$-weight-recovery-secure does not prevent practical attacks that do not rely on full weight recovery.
However, while this security property is too weak to be practically useful for modeling real-world security, this can be useful for proving that some weight release schemes are blatantly insecure by constructing a valid and efficient weight recovery adversary.
In our case study of TaylorMLP, we only used one released weight, but we can also define a stronger variant that allows $\adv$ make multiple queries to $\textsf{Release}(\sk, W)$, which would be convenient for proving insecurity with more than one released weights.

\paragraph{Preventing efficient inference.}
One of the underlying goals for weight recovery attacks is to improve the inference efficiency, as many current weight release schemes are designed to introduce an efficiency gap between the released weight and the original weight.
For example, TaylorMLP aims to slow down the inference of released weight typically up to 8$\times$ to prevent abuse and make a distinction between authorized and free versions of model weight.

Therefore, we introduce a stronger security property which is defined to prevent efficient inference on released model weights.
To measure efficiency without loss of generality, here we consider $\textsf{Cost}(F, x)$ as a function measuring the computational cost of running  $F(x)$ where $F$ is a procedure in our definitions.
$\textsf{Cost}$ can be measured both asymptotic or concrete according to a different context and is often instantiated with the running time of a procedure on a certain input.

We assume that weight release scheme $\Sigma$ is designed such that $\textsf{Cost}(\textsf{Run'}, (\pk, W', x)) \gg \textsf{Cost}(\textsf{Run}, (W, x))$ for $W' \sample \textsf{Release}(\sk, W)$ and $x \in \mathcal{X}$.
Similarly, we introduce the following game for weight release scheme $\Sigma$ on a given training dataset $D \in \mathcal{D}$.
\begin{pcvstack}[center,boxed]
\procedureblock[linenumbering]{$\textbf{Game}_\text{EffiInf}(D, \adv)$}{
W \sample\textsf{Train}(D) \\
(\pk, \sk) \sample \textsf{KGen} () \\
W' \sample \textsf{Release}(\sk, W) \\
x \sample \mathcal{X} \\
y' \sample \textsf{Run'}(\pk, W', x) \\
y^* \sample \adv(\pk, W', x) \\
\pcreturn \textsf{Same}(y^*, y')
}
\end{pcvstack}
The advantage of an adversary $\adv$ in this game is defined as
\begin{align}
    \advantage{\text{EffiInf}}{\Sigma}[(D, \adv)] = \Pr[\textbf{Game}_\text{EffiInf}(D, \adv) \Rightarrow 1]
\end{align}
\begin{definition}
Then for all efficient adversary $\adv$ that runs with computational cost $t$ such that $t < \textsf{Cost}(\textsf{Run'}, (\pk, W', x))$ for $W' \sample \textsf{Release}(\sk, W)$ and $x \in \mathcal{X}$ on some dataset $D \in \mathcal{D}$, if 
\begin{align}
    \advantage{\text{EffiInf}}{\Sigma}[(D, \adv)] \leq \epsilon
\end{align}
for some negligible $\epsilon$ with regard to the security paramter of the scheme $\Sigma$, then the weight release scheme $\Sigma$ is considered to be $(D, t, \epsilon)$-efficient-inference-secure (EffiInf-secure).
\end{definition}

Different from $(D, \epsilon)$-weight-recovery-secure, here we also explicitly consider the computational cost in the security definition.
This is because if $\adv$ is only polynomial time bounded, then $\adv$ can just distill from the released weight into a more efficient weight space $\mathcal{W}^*$.
Therefore, in the definition, we explicitly bound $t < \textsf{Cost}(\textsf{Run'}, (\pk, W', x))$ to ensure a valid adversary with less computation budget than direct inference.
Similarly, we can also define a stronger variant that allows $\adv$ to make multiple queries to $\textsf{Release}(\sk, W)$ and add the query count as a parameter to the security definition.

Note that efficient-inference-secure is a stronger security property than weight-recovery-secure as performing efficient inference does not necessarily need weight recovery.
For example, quantization of released weight can sometimes improve efficiency and thus can be considered a valid attack in this security notion. 
In \S \ref{sec:security-relation} we will prove that efficient-inference-secure implies weight-recovery-secure.

\paragraph{Indistinguishability of released weights.} Inspired by semantic security and IND-CPA properties in cryptographic works\cite{GOLDWASSER1984270}, we can also define a similar security property in the form of indistinguishability, which is shown in the following game for weight release scheme $\Sigma$ on a given training dataset $D \in \mathcal{D}$.
\begin{pcvstack}[center,boxed]
\procedureblock[linenumbering]{$\textbf{Game}_\text{W-IND}(D, \adv)$}{
b \sample \bin \\
(\pk, \sk) \sample \textsf{KGen}() \\
W_0 \sample\textsf{Train}(D) \\
W_1 \sample\textsf{Train}(D) (\neg \textsf{Same}(W_0, W_1))\\
W' \sample \textsf{Release}(\sk, W_{b}) \\\
b' \sample \adv(\pk, W_0, W_1, W') \\
\pcreturn b = b'
}
\end{pcvstack}
The advantage of an adversary $\adv$ in this game is defined as
\begin{align}
    \advantage{\text{W-IND}}{\Sigma}[(D, \adv)] = 2 \left | \Pr[\textbf{Game}_\text{W-IND}(D, \adv) \Rightarrow 1] -\frac{1}{2}\right |
\end{align}
\begin{definition}
Then for all efficient adversary $\adv$ runs within computation cost $t$ on some dataset $D \in \mathcal{D}$, if 
\begin{align}
    \advantage{\text{W-IND}}{\Sigma}[(D, \adv)] \leq \epsilon
\end{align}
for some negligible $\epsilon$ with regard to the security paramter of the scheme $\Sigma$, then the weight release scheme $\Sigma$ is considered to be $(D, t, \epsilon)$-weight-indistinguishability-secure (W-IND-secure).
\end{definition}
Intuitively, a good weight release scheme should leak no information about the original weight, and the game captures this property by letting an adversary distinguish the original weights of a released weight where the original weights are different in value but are trained on the same training data.
If no adversary can win the game with a non-negligible advantage, that means that the scheme does not leak information about the original weight.
Similarly, we can define a stronger variant that allows $\adv$ to make multiple queries to different $(W_0, W_1, W')$ and add the query count as a parameter to the security definition.
By adding conditions on $t$, we can make $(D, t, \epsilon)$-weight-indistinguishability-secure a stronger security property than weight-recovery-secure and efficient-inference-secure.
We will show that weight-indistinguishability-secure implies weight-recovery-secure in \S \ref{sec:security-relation}.

\subsection{Relations of the security properties}
\label{sec:security-relation}
With the previously defined security properties, we show two implication relations between these properties under some assumption that fits the context.
These relations are shown by contraposition, that is, if we want to show $A \rightarrow B$, then it is equivalent to showing $\neg B \rightarrow \neg A$ by constructing an adversary for $A$ from an adversary for $B$.
This is also known as security reductions in cryptographic works.

\paragraph{EffiInf-Secure implies WRec-Secure.} By contraposition, we want to build an adversary $\mathcal{B}$ against EffiInf-Secure game for scheme $\Sigma$ from an adversary $\mathcal{A}$ against WRec-secure game, given that weight release scheme $\Sigma$ satisfies $\textsf{Cost}(\textsf{Run'}, (\pk, W', x)) \gg \textsf{Cost}(\textsf{Run}, (W, x)) + \textsf{Cost}(\mathcal{A}, (\pk, W'))$ for $W' \sample \textsf{Release}(\sk, W)$ and $x \in \mathcal{X}$.
Adversary $\mathcal{B}$ is constructed as follows.
\begin{pcvstack}[center,boxed]
\procedureblock[linenumbering]{$\mathcal{B}_\mathcal{A}(\pk, W', x)$}{
W \sample\mathcal{A}(\pk, W') \\
\pcreturn \textsf{Run}(W, x)
}
\end{pcvstack}
The general idea is to first recover the original weight using $\mathcal{A}$ and then run inference on the recovered weight.
We need to prove that $\mathcal{B}$ is both valid and efficient.
\begin{proof}
We know that $\mathcal{A}$ is valid, meaning that the recovered weight is the same as the original weight.
Then by the correctness of weight release schemes, we have $\textsf{Run}(W, x)$ is the same as $\textsf{Run'}(\pk, W', x)$.
Therefore $\mathcal{B}$ is valid.
By the fact that $\textsf{Cost}(\textsf{Run'}, (\pk, W', x)) \gg \textsf{Cost}(\textsf{Run}, (W, x)) + \textsf{Cost}(\mathcal{A}, (\pk, W'))$ we know that $\textsf{Cost}(\mathcal{B}_\mathcal{A}, (\pk, W', x)) \approx \textsf{Cost}(\mathcal{A}, (\pk, W')) + \textsf{Cost}(\textsf{Run}, (W, x)) \ll \textsf{Cost}(\textsf{Run'}, (\pk, W', x))$, which shows that $\mathcal{B}$ is efficient.
Therefore $\mathcal{B}$ is both valid and efficient.
\end{proof}
From the proof we know that $\mathcal{B}$ wins whenever $\mathcal{A}$ wins, so we have
\begin{align}
    \advantage{\text{WRec}}{\Sigma}[(D, \mathcal{A})] \leq \advantage{\text{EffiInf}}{\Sigma}[(D, \mathcal{B})]
\end{align}
which gives EffiInf-Secure implies WRec-Secure.

\paragraph{W-IND-Secure implies WRec-Secure.} Similarly by contraposition, we want to build an adversary $\mathcal{B}$ against W-IND-Secure game for scheme $\Sigma$ from an efficient adversary $\mathcal{A}$ against WRec-Secure game under some feasible assumptions.
Adversary $\mathcal{B}$ is constructed as follows.
\begin{pcvstack}[center,boxed]
\procedureblock[linenumbering]{$\mathcal{B}_\mathcal{A}(\pk, W_0, W_1, W')$}{
W \sample\mathcal{A}(\pk, W') \\
\pcif \textsf{Same}(W, W_0) \pcthen \pcreturn 0\\
\pcelse \pcreturn 1
}
\end{pcvstack}
The general idea is to first recover the original weight using $\mathcal{A}$ and then distinguish the original weight.
Similarly, we need to prove that $\mathcal{B}$ is both valid and efficient.
\begin{proof}
We know that $\mathcal{A}$ is valid, meaning that the recovered weight is the same as the original weight.
Because $\textbf{Game}_\text{W-IND}$ ensures that $\neg \textsf{Same}(W_0, W_1)$, so we have either $\textsf{Same}(W_0, W)$ or $\textsf{Same}(W_1, W)$.
$\mathcal{B}$ runs with constant extra steps compared with $\mathcal{A}$.
Since $\mathcal{A}$ is efficient, $\mathcal{B}$ is also efficient.
Therefore $\mathcal{B}$ is both valid and efficient.
\end{proof}
From the proof we know that $\mathcal{B}$ wins whenever $\mathcal{A}$ wins, so we have
\begin{align}
    \advantage{\text{WRec}}{\Sigma}[(D, \mathcal{A})] \leq \advantage{\text{W-IND}}{\Sigma}[(D, \mathcal{B})]
\end{align}
which gives W-IND-Secure implies WRec-Secure.


%% file: parts/4-TaylerAttack.tex
\section{Case Study: Insecurity of TaylorMLP}
\label{sec:talyerattack}

To show how our security definition can be applied to the analysis of real-world weight release schemes, we present a case study on a recent scheme named TaylorMLP.
In this case study we will have a brief review of how TaylorMLP works and then we present an attack against the weight-recovery-security of TaylorMLP with experimental results.
We then discuss the implications of this attack.

\begin{table*}[htb]
    \centering
    \begin{tabular}{
        lcccccc
    }
        \toprule
         Model & OPT-125M & OPT-1.3B & OPT-2.7B & OPT-6.7B & Llama2-7B & Llama2-13B \\
         \midrule
         \# TaylorMLP Parameters & 28,311,552  & 402,653,184  & 838,860,800  & 2,147,483,648  & 1,442,840,576 & 2,831,155,200 \\
         \# Recovered Parameters & 28,310,784  & 401,649,665  & 837,347,840  & 2,144,915,459 & 1,442,840,576 & 2,831,155,200 \\
         Recovered Ratio & 99.99\%  & 99.76\%  & 99.82\%  & 99.88\%  & 100.00\% &100.00 \% \\
         \midrule
         Running Time & 19.70 s & 116.79 s & 187.88 s & 417.92 s & 318.60 s & 571.05 s \\
         Attack Cost in USD & 0.01 & 0.05 & 0.07 & 0.17 & 0.13 & 0.22 \\
        \bottomrule
    \end{tabular}
    \caption{
        Attack Performance. ``\# TaylorMLP Parameters" refers to the total number of weight parameters processed by TaylorMLP in the model, while ``\# Recovered Parameters" denotes the number of weight parameters successfully recovered by our attack. ``Recovered Ratio" is defined as the proportion of weights that can be successfully recovered using our proposed method, whereas ``successfully" is defined as the relative error of the recovered weights is less than 1\%. Relative error is calculated as ${|\mathbf{W}_{\text{rec}} - \mathbf{W}|}/{|\mathbf{W}|} \cdot100\%$ where $\mathbf{W}_{\text{rec}}$ denotes the recovered weights and $\mathbf{W}$ denotes the ground-truth weights.
    }
    \label{tab:res-attack-results}
\end{table*}

\subsection{TaylorMLP}
Based on syntax of weight release schemes described in \S\ref{sec:security-syntax}, TaylorMLP can be seen as a weight release scheme $\Sigma = (\textsf{Train}, \textsf{Run}, \textsf{KGen}, \textsf{Release}, \textsf{Run'})$, where
\begin{itemize}
    \item $\textsf{Train}$ is the original LLM training process on the training dataset.
    
    \item $\textsf{Run}$ is the original LLM inference algorithm on the input prompt using the original model weight.
    
    \item $\textsf{KGen}$ is not instantiated with a meaningful algorithm, and in this case $\mathcal{K}_\sk $ and $\mathcal{K}_\pk$ are both $\emptyset$.
    
    \item $\textsf{Release}$ is the algorithm that converts MLP in LLM weight into TaylorMLP format as is shown in Algorithm \ref{alg:Taylor-mlp}. 
    
    \item $\textsf{Run'}$ is a LLM inference algorithm on input prompt using TaylorMLP format weight.
\end{itemize}

\begin{algorithm}[tb]
   \caption{Transforming MLP to TaylorMLP}
   \label{alg:Taylor-mlp}
\begin{algorithmic}[1]
\REQUIRE MLP($\bullet | \mathbf{V}, \mathbf{b}, \mathbf{W}, \mathbf{c}$) and $\mathbf{z}_0$
\ENSURE TaylorMLP($\bullet | \mathbf{V}, \mathbf{z}_0, \{\Theta_{i,0}, \cdots, \Theta_{i,N}\}_{i=1}^D$)
\FOR{$i := 1$ to $D$}
    \STATE $\mathbf{W}_i$ and $c_i$ take the $i$-th row and $i$-th element of $\mathbf{W}$ and $\mathbf{c}$, respectively.
    \STATE $\Theta_{i,0} = \mathbf{W}_i \odot \text{Act}(\mathbf{z}_0 + \mathbf{b}) + c_i$
    \FOR{$n := 1$ to $N$}
        \STATE $\Theta_{i,n} = \mathbf{W}_i \odot \text{Act}^{(n)}(\mathbf{z}_0 + \mathbf{b})(n!)^{-1}$
    \ENDFOR
\ENDFOR
\end{algorithmic}
\end{algorithm}
In short, TaylorMLP transforms $(\textbf{b},\mathbf{W}_i,c_i)$ into $(\textbf{z}_0,[\Theta_{i,0},\Theta_{i,1}, ... , \Theta_{i,N}])$ by calculating
\begin{align}
    \textbf{z}_0 &= \frac{\textbf{z}_{\max} + \textbf{z}_{\min}}{2} \\
    \Theta_{i,n} &= \mathbf{W}_i \odot \frac{\text{Act}^{(n)}(\textbf{z}_0+\textbf{b})}{n!} \label{eq13}
\end{align}
where $\text{Act}^{(n)}$ is the $n$-th order derivative of activation function and $\textbf{z}_{\max / \min}$ is the $\max / \min$ value collected on some test input.
The correctness of TaylorMLP can be shown below.
\begin{align}
    y_i &= \text{Act}(\textbf{z}+\textbf{b}) \cdot \mathbf{W}_i + c_i \\
    &\approx \left\langle \mathbf{W}_i, \sum_{n=0}^{N} \frac{\text{Act}^{(n)}(\mathbf{z}_0 + \mathbf{b})}{n!} \odot (\mathbf{z} - \mathbf{z}_0)^n \right\rangle + c_i \\
    &= \sum_{n=0}^{N} \left\langle \mathbf{W}_i \odot \text{Act}^{(n)}(\mathbf{z}_0 + \mathbf{b})(n!)^{-1}, (\mathbf{z} - \mathbf{z}_0)^n \right\rangle\\
    &= \sum_{n=0}^{N} \left \langle \Theta_{i,n}, (\textbf{z}-\textbf{z}_0)^n \right \rangle 
\end{align}

\subsection{The weight recovery attack}
The general idea is that while TaylorMLP introduces randomness in $\textbf{z}_0$, since most transformations are elementwise and more parameters are added, it is possible to solve equations from the released weight to recover the original weight.
We present our attack for weight recovery as follows.
\paragraph{The attack.} We start by transforming Equation \ref{eq13}, which gives
\begin{align}
    \mathbf{W}_i = (n!)\Theta_{i,n} \odot \frac{1}{\text{Act}^{(n)}(\mathbf{z}_0 + \mathbf{b})}
\end{align}
We can see that as long as we can solve for the value of $\mathbf{b}$, we can solve $\mathbf{W}_i$ given $\text{Act}^{(n)}$ is invertible in some range.
Note that we have multiple such equations. Consider a pair $(a, b)$ satisfying $0 \leq a < b \leq N$, we have
\begin{align}
    \mathbf{W}_i &= (a!)\Theta_{i,a} \odot \frac{1}{\text{Act}^{(a)}(\mathbf{z}_0 + \mathbf{b})}\\
    &= (b!)\Theta_{i,b} \odot \frac{1}{\text{Act}^{(b)}(\mathbf{z}_0 + \mathbf{b})}
\end{align}
After rearranging, we get
\begin{align}
    \frac{(a!)\Theta_{i,a}}{(b!)\Theta_{i,b}} = \frac{\text{Act}^{(a)}(\mathbf{z}_0 + \mathbf{b})}{\text{Act}^{(b)}(\mathbf{z}_0 + \mathbf{b})}
\end{align}
Since this equation holds elementwise, for each element of $(\mathbf{z}_0 + \mathbf{b})$, we can obtain one equation.
To simplify the notation, we denote $ f_{a,b}(x) = \frac{\text{Act}^{(a)}(x)}{\text{Act}^{(b)}(x)}$.
Then we have
\begin{align}
    \frac{(a!)\Theta_{i,a}[j]}{(b!)\Theta_{i,b}[j]} = f_{a, b}((\mathbf{z}_0 + \mathbf{b})[j])
\end{align}
for $j \in [D]$.
Considering that for most pairs $(a, b)$, $f_{a,b}(x)$ is expected to be invertible, we can recover $(\mathbf{z}_0 + \mathbf{b})$ with high probability, thus solving for the value of $\mathbf{b}$, and consequently recovering $\mathbf{W}_i$.

\paragraph{Efficiency.} The attack is efficient given $\text{Act}$ is efficiently invertible using numerical methods like Newton's method. 
For activation functions like SiLU and GeLU as used in the original TaylorMLP paper, the attack takes $O(N\cdot \#\text{Params})$ time and space, which is efficient.

\paragraph{Numerical stability.} In the real scenario for activation functions like SiLU, $f_{a,b}(x) = \frac{\text{Act}^{(a)}(x)}{\text{Act}^{(b)}(x)}$ suffers from occasional problems with floating point numerical stability on some outlier weight values.
This is partly due to the widespread use of float16 in LLM inference and switching to double-precision floating point numbers mitigates this issue.
However, due to the fractional nature of $f_{a,b}$, some outlier weight values can not be reliably solved due to numerical stability issues, and this leads to a small portion ($<1\%$) of weights that can not be recovered in practice, which is discussed in our experiments.

\subsection{Experiments}

\paragraph{Experiment settings.} Following the evaluation in the original TaylorMLP paper, we choose OPT and Llama model family to evaluate the effectiveness of our attack.
We process the MLP layers of each transformer block of the model weights to get the released weight.
We then implemented our attack in Python 3 using numpy and scipy.
We run the attack to recover the MLP layer weights using the released weight as input.
All experiments were conducted on an x86 Linux machine equipped with 20 CPU cores and 128 GB of memory without GPU acceleration.

We measure the number and ratio of successfully recovered weight values as well as running time and estimated cost of the attack.
A weight value is considered recovered successfully if its error compared with the original value is less than $1\%$.
Estimated cost is calculated based on the price of instances from mainstream cloud providers that are comparable to the x86 Linux machine used in the experiment.

\paragraph{Results.} The results are shown in Table \ref{tab:res-attack-results}. 
In our experiments, models from OPT and Llama of various sizes consistently achieved recovered rates very close to 100\%, indicating that the relative error between $\mathbf{W}_{\text{rec}}$ and $\mathbf{W}$ is negligible.
Furthermore, due to differences in architecture and weight distribution between the Llama-2 and OPT models, the relative error also varies.
Specifically, the recovery results for Llama-2 are noticeably better than those for OPT.

We measured the time required to recover the entire set of model weights.
As shown in our results, the OPT-125M model requires less than 20 seconds, while the largest model tested, Llama-2-13B, takes under 10 minutes.
The recovery process is quite efficient, even on a standard CPU machine.

We then roughly estimated the monetary cost associated with the weight recovery process.
Referring to \texttt{m8g.8xlarge} instance with a similar configuration on AWS, the price is approximately \$1.43616 per hour. 
Therefore the highest cost among tested models is only \$0.22 for a successful attack for computation.
This demonstrates that the attack is cost-effective and feasible even for attackers with limited budget in practice.

\begin{figure}[htbp]
    \centering
    \includegraphics[width=0.5\textwidth]{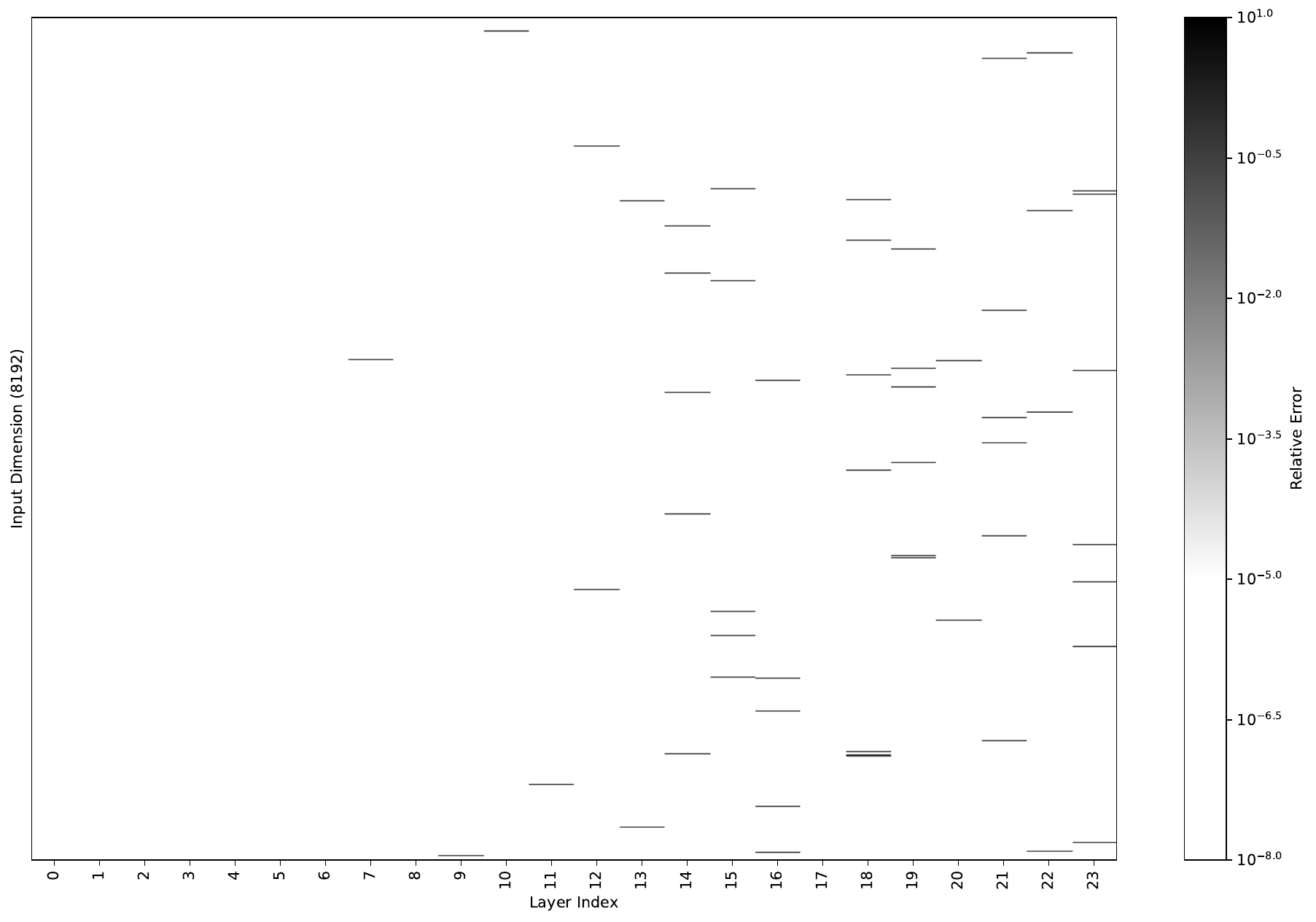}
    \vspace{-0.7cm}
    \caption{Relative Error for recovered weights in OPT-1.3B. We concatenate weights across all layers of the OPT-1.3B model along the x-axis for comprehensive visualization. To enhance visual clarity and emphasize variations, we apply a $\log_{10}$ scale to the relative error values.}
    \label{fig:heatmap}
\end{figure}

Additionally, we visualize the relative error between the ground-truth weights and the recovered weights using a heat map. 
As shown in Figure~\ref{fig:heatmap}, the majority of errors are extremely low, with only a few channels exhibiting relatively larger errors.
This demonstrates the effectiveness of our recovery method, as it is able to accurately recover most of the elements.

\subsection{Takeaways}
The attack demonstrates that TaylorMLP is vulnerable to our weight recovery attack, thereby failing to provide weight-recovery-security. From the security relations established in \S\ref{sec:security-relation}, TaylorMLP consequently also fails to ensure efficient-inference-security and weight-indistinguishability-security. These results establish the insecurity of TaylorMLP as a weight release scheme designed to protect model ownership and prevent unauthorized use.

While the original TaylorMLP paper conducted detailed evaluations of correctness and efficiency gaps, and claimed that fine-tuning the released model is infeasible, the authors did not formulate formal security definitions or provide rigorous security proofs under reasonable assumptions. This lack of formal security analysis led to unnoticed design vulnerabilities that we successfully exploited in our attack.

%% file: parts/5-FutureWork.tex
\section{Design Principles for Future Weight Release Schemes}
\label{sec:futurework}
Based on our security definitions and analysis of TaylorMLP, we outline key principles for designing and evaluating future weight release schemes.

\paragraph{Clear security goals.} Current real-world weight release schemes are usually exclusive, meaning that two weight release schemes are generally not easily combined to hedge the risk. For example, if a developer decides to protect the model using TaylorMLP and has released the model in this format, he or she may not easily switch to another scheme given TaylorMLP is broken, thus the already released model is now under threat. 

Depending on typical use cases for developers, users and adversaries, security goals can vary. While informal security notions are not enough against real-world attacks, they are a good starting point for sketching the security goals of a weight release scheme. Future schemes should begin with precise security definitions that specify the adversarial model, threat capabilities, and desired security properties, drawing from established cryptographic frameworks such as digital signature schemes\cite{guo2023sok}.

\paragraph{Provable security.} To ensure robust security guarantees, weight release schemes should provide formal security proofs. 
A security proof requires: 1) formal syntax defining the scheme's algorithms, 2) explicit computational hardness assumptions, 3) precise security definitions, and 4) rigorous proofs using established techniques such as security reductions. 
Such proofs relate the scheme's security to widely accepted hardness assumptions, providing theoretical foundations beyond informal arguments.

\paragraph{Offensive evaluation.} Provable security alone is insufficient due to gaps between theoretical models and real-world implementations\cite{koblitz2007another}. 
Schemes require thorough offensive evaluation to identify potential vulnerabilities before deployment. 
As demonstrated in our TaylorMLP analysis, successful attacks serve as ``proofs of insecurity" that can invalidate informal security claims. 
This adversarial testing reveals practical limits and guides parameter selection for secure deployment.

\paragraph{Compatibility requirements.} Beyond security, practical adoption requires compatibility with existing ML infrastructure. 
For example, the $\textsf{Run'}$ algorithm may integrate seamlessly with mainstream inference frameworks to avoid imposing additional implementation burdens on users. 
Schemes that require specialized execution environments or exotic data formats can face significant adoption barriers regardless of their security properties.

%% file: parts/6-Conclusion.tex
\section{Conclusion}
\label{sec:conclusion}

We have taken a step toward establishing a rigorous theoretical foundation for secure weight release by formalizing security definitions for weight release schemes.
Our case study of TaylorMLP reveals model weight recovery vulnerabilities, demonstrating that existing schemes fail to achieve their informal security claims and highlighting the gap between promised and actual security guarantees.

We hope this work will advocate for rigorous research at the intersection of machine learning and security communities. 
By providing concrete security definitions and demonstrating their application, we establish a blueprint for designing and evaluating future secured weight release schemes that provide meaningful security guarantees rather than false assurances.

%% file: parts/7-Appendix.tex
\onecolumn
\section{Analysis of numerical stability of SiLU/GeLU derivative.}

\begin{figure*}[tbp]
    \centering
    \includegraphics[width=1.0\textwidth]{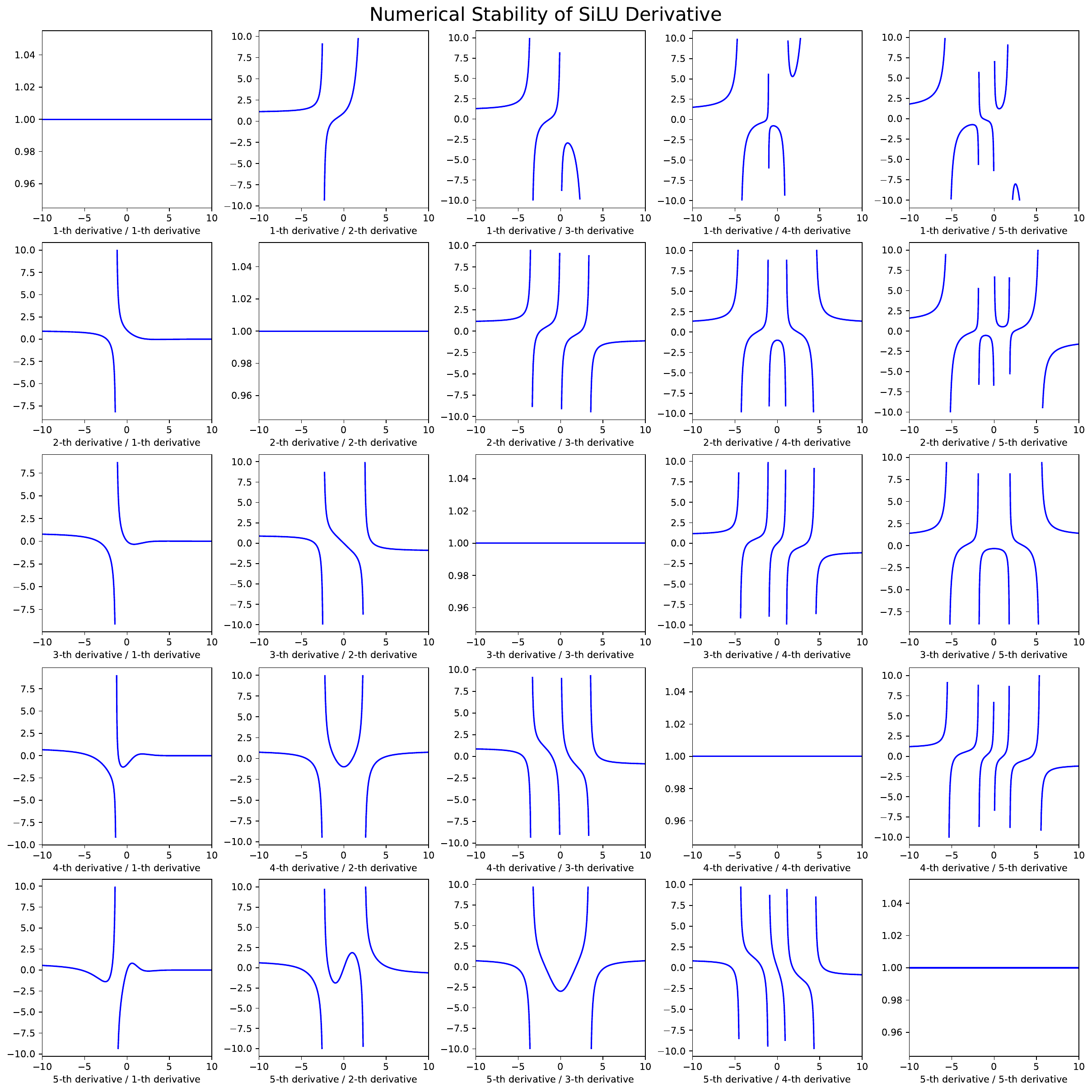}
    \caption{Numerical stability analysis of SiLU derivative ratios. Each subplot shows the ratio between SiLU derivatives from first to fifth order, evaluated over the interval. Diagonal entries represent ratios of derivatives with themselves and therefore equal 1. Many off-diagonal entries exhibit significant numerical instability, characterized by large fluctuations or extreme values, particularly away from zero. These instabilities highlight potential sources of errors and inaccuracies in practical computations involving higher-order SiLU derivatives.}
    \label{fig:silu}
\end{figure*}

\begin{figure*}[tbp]
    \centering
    \includegraphics[width=1.0\textwidth]{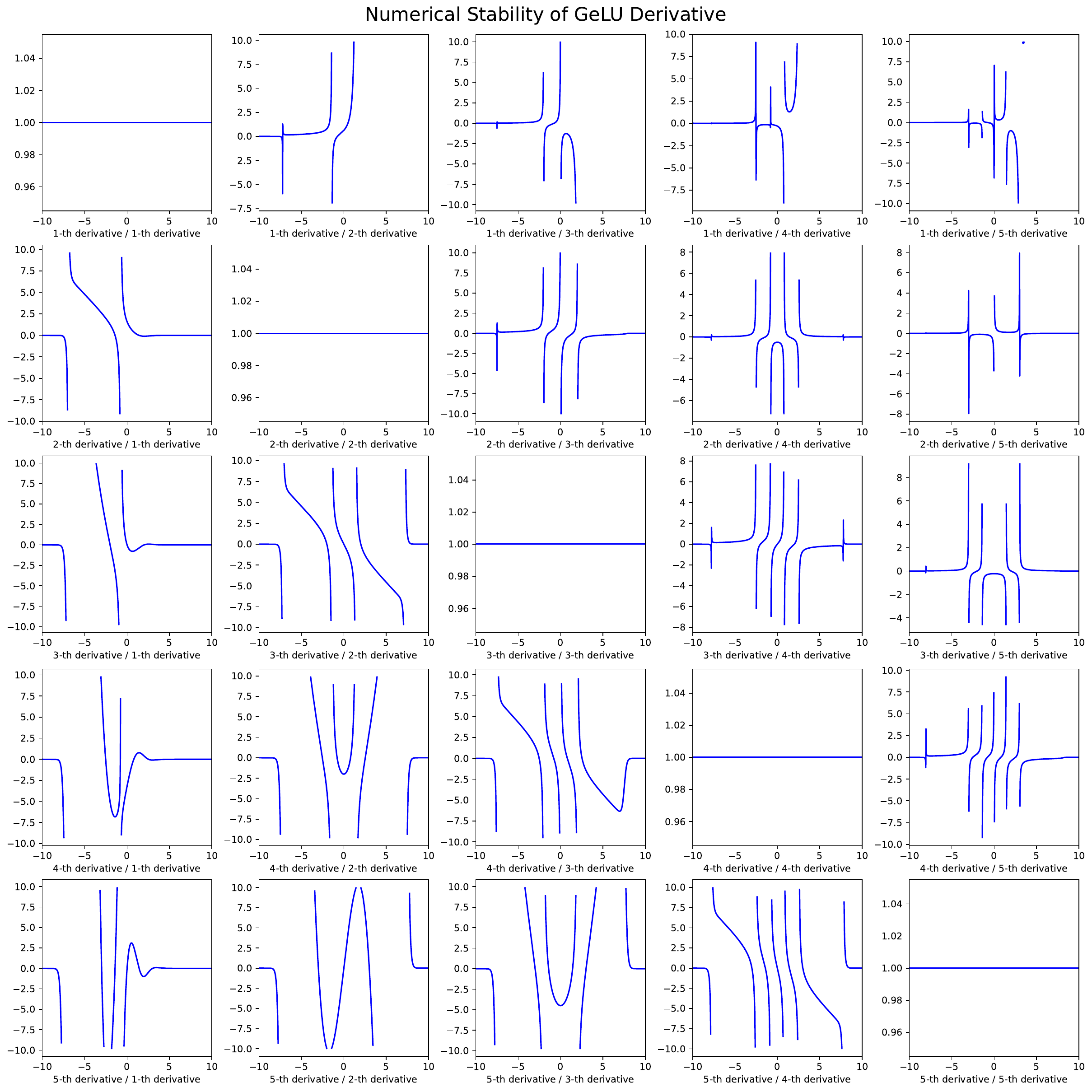}
    \caption{Numerical stability analysis of GeLU derivative ratios. Similar to SiLU, ratios between GeLU derivatives from first to fifth order show numerical instabilities characterized by large fluctuations and extreme values, particularly away from zero. Although minor differences exist, overall stability patterns closely resemble those observed for SiLU.}
    \label{fig:gelu}
\end{figure*}

The numerical stability of higher-order derivatives of the SiLU and GeLU activation function exhibits significant sensitivity when evaluating ratios of derivatives. 
As demonstrated in Fig.\ref{fig:silu} and Fig.\ref{fig:gelu}, ratios of SiLU derivatives within the interval \([-10, 10]\) frequently encounter numerical instability, often resulting in large fluctuations or undefined values at various points away from zero. 
While stability improves in the vicinity of zero due to more balanced numerical magnitudes, even slight deviations or outliers in the input distributions can cause pronounced instabilities. 
Importantly, since different network architectures and model sizes inherently produce distinct internal numerical distributions, the resulting numerical instabilities manifest differently across these models, leading to varying degrees of reconstruction errors and inaccuracies. 
Therefore, the numerical instability induced by varying distributions across different model architectures and scales inevitably introduces a certain degree of error into our reconstruction method, underscoring the necessity for careful numerical considerations when employing higher-order SiLU derivatives in analysis and optimization tasks.

\section{Computational Infrastructure.}
Table \ref{tab:setting} provides details on the computational infrastructure and environment information.

\begin{table}[H]
\centering
\begin{tabular}{l|l}
\toprule
\textbf{Name} & \textbf{Value} \\
\midrule
CPU  &    Intel Core Ultra 7 265K      \\ 
Memory     &     128GB      \\ 
\midrule
Data type     &    torch.bfloat16       \\ 
\midrule
OS          &   Debian GNU/Linux trixie/sid \\
\midrule
Python      &   3.13.3  \\
numpy      &   2.2.6  \\
torch           &   2.7.1   \\
transformers      &   4.52.4  \\
scipy           &   1.15.3   \\
matplotlib      &   3.10.3  \\
\bottomrule
\end{tabular}
\caption{
Configuration of experiment and computing infrastructure.
    }
    \label{tab:setting}
\end{table}

